\documentclass[conference]{IEEEtran}
\IEEEoverridecommandlockouts
 \usepackage{lipsum,graphicx,multicol}
\usepackage{color,soul}
\usepackage{mathtools}
\usepackage{graphicx}
\usepackage{color,soul}
\usepackage{setspace}
\usepackage{multirow}
\usepackage{tablefootnote}
\usepackage[norelsize, linesnumbered, ruled, lined, boxed, commentsnumbered]{algorithm2e}
\usepackage[normalem]{ulem}
\usepackage[font=small,labelfont=bf]{caption}
\usepackage{subcaption}

\usepackage{bm, amsmath, amssymb, amsthm}
\usepackage{amsfonts}
\usepackage {amssymb,amsmath,xcolor}
\usepackage{booktabs}
\usepackage{cite}
\usepackage{array}
\usepackage[acronym]{glossaries}
\usepackage{cases}
\usepackage{cleveref}
\usepackage{amsthm,mdframed,calc}

\def\BibTeX{{\rm B\kern-.05em{\sc i\kern-.025em b}\kern-.08em
    T\kern-.1667em\lower.7ex\hbox{E}\kern-.125emX}}
    
  \newtheorem{theorem}{Theorem}

\newtheorem{problem}{Problem}

\newcommand{\bieee}{\begin{eqnarray}{rCl}}
\newcommand{\eieee}{\end{eqnarray}}

\SetKwInput{KwInput}{Input}                
\SetKwInput{KwOutput}{Output}              

\begin{document}

\title{Network-Centric Countermeasures Against Integrated Sensing Enabled Jamming Adversaries
}

\author{\IEEEauthorblockN{Soumita Hazra and J. Harshan }
\IEEEauthorblockA{{Indian Institute of Technology Delhi, India}}
}

\maketitle

\begin{abstract}
Recent developments in Integrated Sensing and Communication have led to new adversarial models in wireless security through Integrated Sensing and Jamming (ISAJ) adversaries. ISAJ adversaries, owing to their sensing capabilities, are known to inject jamming energy over the victim's frequency band, and also use generalized energy measurements on various network frequencies to detect the presence of countermeasures. Existing countermeasures against such ISAJ adversaries are laid under the assumption that the adversary does not have the knowledge of the countermeasure. However, according to Kerchoffs' principle in cryptography, security of a countermeasure should only rely on the secret-keys, not on the obfuscation of the countermeasure. On testing the security of existing countermeasures, we observe that they violate Kerchoffs' principle, thus motivating the need for new countermeasures. In this regard, we propose a novel network-centric countermeasure against ISAJ adversaries, wherein a group of users in the network assist the victim to reliably communicate her messages in a covert manner. Firstly, we analyse the error performance of the proposed countermeasure, and study its behavior on the number of assisting users in the network. Subsequently, to validate its security against Kerchoffs' principle, we study the Shannon's entropy associated with the presence of the victim's messages in the network and analyse its behaviour as a function of the number of assisting users. Finally, to study the interplay between reliability and covertness, we pose interesting optimization problems and solve them to choose the underlying parameters of the countermeasure and the number of assisting users. 
\end{abstract}


\section{Introduction}

Integrated Sensing and Communication (ISAC) \cite{ISAC1,ISAC2} has emerged as a promising technology for $6$G wireless communication networks, owing to its capability to merge sensing and communication in a single system. This integration has not only improved spectrum efficiency and reduced hardware costs, but has enabled wireless networks to act as sensors. However, in the parallel world of wireless security, recent developments in ISAC has also led to new threat models involving jamming adversaries that are enabled with integrated sensing capabilities. Such a class of adversaries, henceforth referred to as Integrated Sensing and Jamming (ISAJ) adversaries \cite{FDJ2,V1,NCC,TVT1}, intend to execute denial-of-service (DoS) \cite{DOS1,Jamming2022} attack on victim's messages, by injecting jamming energy on the band, akin to traditional jammers. Furthermore, to enable the integrated sensing capabilities, ISAJ adversaries are also equipped with full-duplex radios (FDR) \cite{SIC2,FDR1,LInew1} to monitor different network frequency bands to detect the presence of potential countermeasures, while simultaneously injecting jamming energy over the victim's frequency band. The ISAJ adversary proposed in \cite{V1} monitors the average energy of the received symbols over the victim's frequency band to detect the presence of countermeasures. To evade this attack, the countermeasure proposed in \cite{V1} ensures that the average energy of the received symbols over victim's frequency band is consistent, before and after the countermeasure with a high probability.
Further, \cite{NCC,TVT1} considered a stronger ISAJ adversary, which monitors the instantaneous energy of the received symbols over all the network frequencies to detect the presence of countermeasure.
Also, the ISAJ adversary compares the statistical distribution of the received symbols before and after the attack using Kullback-Leibler divergence based detector.
To tackle this ISAJ adversary, the countermeasures proposed in \cite{NCC,TVT1} ensure that the instantaneous energy of the received symbols before and after  the countermeasure is consistent with a high probability, along with maintaining 
the statistical distribution of the received symbols over all the network frequencies. 

\subsection{Motivation}
We highlight that the countermeasures proposed in  \cite{V1,NCC,TVT1} are laid under the assumption that 
the ISAJ adversary monitors the statistics of the network frequencies without the knowledge of the countermeasure.
However, according to Kerchoffs' principle in cryptography \cite{KP1}, for a countermeasure to be secure, details of the countermeasure should be made public, except the secret-keys. 
In other words, security of a countermeasure should not rely on the obfuscation of the countermeasure, rather it should rely only on the secret-keys.
As a result, it is imperative to test if the countermeasures in \cite{NCC,TVT1} remain secure when their details are made public. Towards that direction, we recall that in \cite{NCC,TVT1}, the victim is asked to move to another frequency band that is unknown to the ISAJ adversary, therein the victim and the helper, cooperatively transmit their information symbols to the destination using a new modulation technique. 
When the countermeasure is made public, the ISAJ adversary knows that the victim has moved to some other frequency, and is using a specific modulation technique to communicate her messages. However, the ISAJ adversary does not know the frequency band of the helper node. As a result, the uncertainty, which is quantified by Shannon's residual entropy \cite{SE1}, associated with the helper's frequency band before observing the network frequencies is $\mbox{log}_{2}(L-1)$ bits, where $L$ is the total number of frequency bands. For the countermeasures in \cite{NCC,TVT1} to be secure as per Kerchoffs' principle, the amount of uncertainty associated with the helper's frequency band should be close to $\mbox{log}_{2}(L-1)$ bits even after observing all the network frequencies. However, we observe that when the countermeasures in \cite{NCC,TVT1} are implemented, the ISAJ adversary can tune into all the network frequencies to detect the specific modulation technique of the countermeasure. Owing to the fact that there is no change in the modulation techniques of the other users in the network, this further implies that the residual entropy associated with the helper's frequency band will be small. As a consequence, the countermeasures in \cite{NCC,TVT1} are not secure against Kerchoffs' principle.

\subsection{Contribution}
For the ISAJ adversary with the capabilities mentioned in  \cite{NCC,TVT1}, we introduce a new threat model abiding Kerchoffs' principle wherein all the details of the prospective countermeasure should be made public, except the secret-keys. Against this threat model, we propose a novel countermeasure referred to as the Network-Centric Mitigation Strategy (NCMS), wherein, we ask the victim to move to a helper's frequency band. Subsequently, the victim and the helper cooperatively transmit their information symbols to the destination using a portion of their energies, while they pour their remaining energies over victim's band using a shared secret-key.
In addition to this, to address Kerchoffs' principle, we ask some of the network's users to work in groups, and mimic the signalling waveforms over the helper's band. In particular, the users in a group, use a part of their energies to communicate their information to the destination, and use the remaining energies to mimic the transmission over the helper's band.
In NCMS since the network users deviate from their regular signalling waveforms, their error performance will be impacted. As a result, we analyse the error performance of NCMS as a function of the number of users mimicking the transmission over the helper's band and the manner in which they share their energies. Subsequently, to validate the security strength as per Kerchoffs' principle, we study the covertness of the NCMS as a function of the number of users mimicking the transmission over the helper's band. First, we show that the measured residual entropy associated with the helper's band is close to the ideal entropy values, which is governed by the number of users mimicking the helper's frequency band.
Therefore, to jointly achieve reliable and covert communication under such ISAJ adversaries, we solve optimization problems to find the optimal number of users that should mimic the transmissions over the helper's band, and the optimal value of energy usage for cooperation. Extensive analysis backed by thorough simulation results reveal that reliable communication can be achieved even under the stringent Kerchoffs' principle, however with acceptable degradation in the error performance compared to the baselines that violate Kerchoffs' principle.

\section{System Model}

\begin{figure}[h]
\begin{center}
\includegraphics[scale = 0.31]{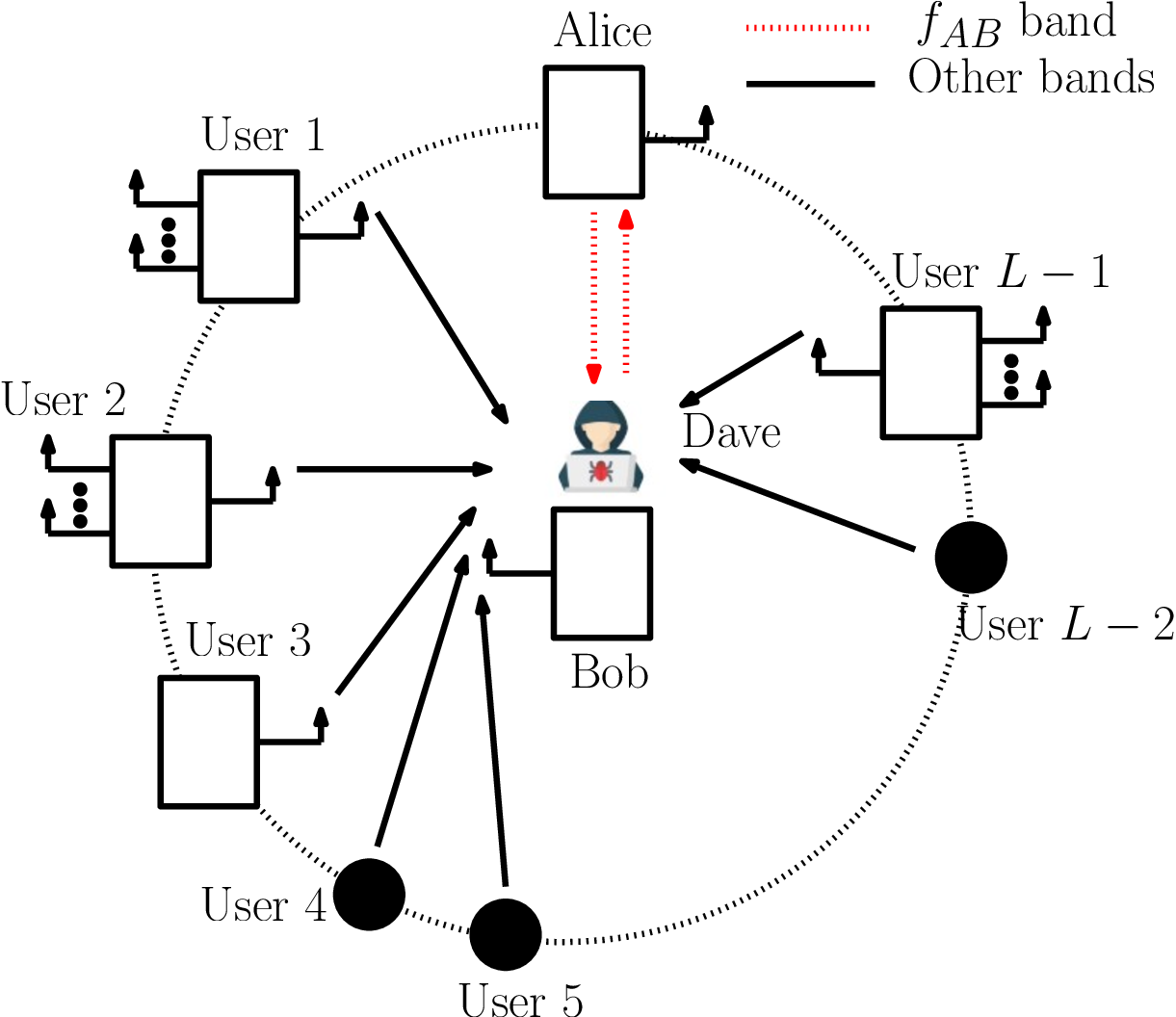}
\end{center}
\vspace{-0.12cm}
\caption{Depiction of the network model, wherein  Alice and the rest of the $L-1$ network users communicate with Bob, over their allocated uplink frequencies. In this network, the ISAJ adversary, namely Dave, injects jamming energy over Alice's frequency band, while monitoring different network frequencies to detect countermeasures.}
\label{SMdiag}
\end{figure}

We consider a heterogeneous wireless network consisting of  $L$ users,  as shown in Fig. \ref{SMdiag}, which communicate with a base-station, namely Bob, using their uplink channels on orthogonal frequencies. Also, all the uplink frequencies provided to Bob are assigned to the users of the network, and as a result, the network is crowded. Additionally, some of these $L$ users have the requirement of high-spectral efficiency on their uplink frequencies, as a result, they are equipped with an FDR with multiple-receive antennas and single-transmit antenna. However, there is no such requirement at the other users, consequently, they are not equipped with an FDR. Of these $L$ users, one user, namely Alice, has critical information to communicate with Bob, over the frequency band, denoted as $f_{AB}$.
Therefore, she is a potential victim of DoS attack from an ISAJ adversary. Owing to the fact that coherent modulation schemes are susceptible to pilot contamination attack from an ISAJ adversary, Alice communicates her information modulated using non-coherent On-Off keying (OOK). The rest of the $L-1$ users are not potential victims of DoS attack from an ISAJ adversary, as a result, they use a coherent modulation scheme, to communicate their information to Bob.

Similar to \cite{NCC,TVT1}, we assume the presence of an ISAJ adversary, namely Dave, 
who execute DoS attack on the critical information of Alice.
To achieve this, Dave injects jamming energy on the $f_{AB}$ band, and he is also placed close to Bob so that the impact of the jamming energy is maximised.
In addition to this, Dave is equipped with an FDR, as a result, he can detect the presence of potential countermeasures while injecting jamming energy on the $f_{AB}$ band. We also assume that Dave can tune into all the network frequencies and observe the received symbols. While our threat model is similar to that in \cite{NCC,TVT1}, our main departure is the incorporation of Kerchoffs' principle, which assumes that all the details of the countermeasure is available to Dave, except the secret-keys.

\subsection{Impact of using Kerchoffs' principle on \cite{NCC,TVT1}}

Recall that \cite{NCC,TVT1} ask Alice to move to another frequency band, wherein Alice and the helper, cooperatively transmit their information symbols to Bob, by using a new modulation technique. Since only one out of the $L-1$ frequency bands undergoes changes in the modulation technique, Dave can detect this change with high probability using the knowledge of the modulation. As a consequence, he can generate a posteriori probability values on each frequency band on the likelihood of being the helper's band, and then measure the associated Shannon's residual entropy. For instance, upon implementing one such method against the countermeasure in \cite{NCC}, we observe that the normalised residual entropy values (measured entropy values normalized by $\mbox{log}_{2}(L-1))$ are upper bounded by 0.5 for $L = 42$ and SNR values of 20 dB, 25 dB, 30 dB and 35 dB. However, ideal normalised entropy is expected to be close to one, and thus we conclude that the countermeasure proposed in \cite{NCC} is not secure against Kerchoffs' principle.
On this note, we ask: \textit{How to design network-centric countermeasures that are reliable and also ensure that the Shannon's residual entropy on the victim's messages in the network is within an acceptable range?}

\section{Network-Centric Mitigation Strategy} \label{CH}
In this section, we propose a countermeasure called the Network-Centric Mitigation Strategy (NCMS), wherein 
we ask Alice to tune into the frequency band of a helper node in the network, referred to as Charlie, to reliably communicate her information to Bob.
Further, Alice and Charlie use a part of their energies to communicate their information to Bob, and they pour their residual energies over the $f_{AB}$ band using a shared secret-key.
Assuming that Charlie, is equipped with an FDR of $N_C$ receive-antennas and single-transmit antenna, and is communicating on the $f_{CB}$ band, we present our strategy on the $f_{CB}$ band and the $f_{AB}$ band, in Section \ref{fcbtheory} and Section \ref{fabtheory}, respectively. 
Now, to increase the entropy at Dave regarding the helper's frequency band, we ask $L_C$ users of the network, for $L_C=2c$, $c \in \mathbb{N}$, to reorganise themselves into groups of two users each, so as to mimic the signalling scheme over the $f_{CB}$ band. 
These $L_C$ users will communicate in their usual frequency band using a part of their energy, and they will pour their remaining energy into their partner's frequency band using a shared secret-key, in such a way that the signalling scheme in each of the $L_C$ bands  will mimic the signalling scheme over the $f_{CB}$ band. 
For description, we assume that one such pair is formed by users namely, Tom and Henry. As the signalling scheme in their frequency bands have same structure, we will only explain the strategy over Henry's frequency band, denoted as $f_{HB}$ band, in Section \ref{fhbtheory}.

\subsection{Strategy over the $f_{CB}$ band}\label{fcbtheory}
\begin{figure*}
     \subfloat[]
{
\includegraphics[scale=0.27]{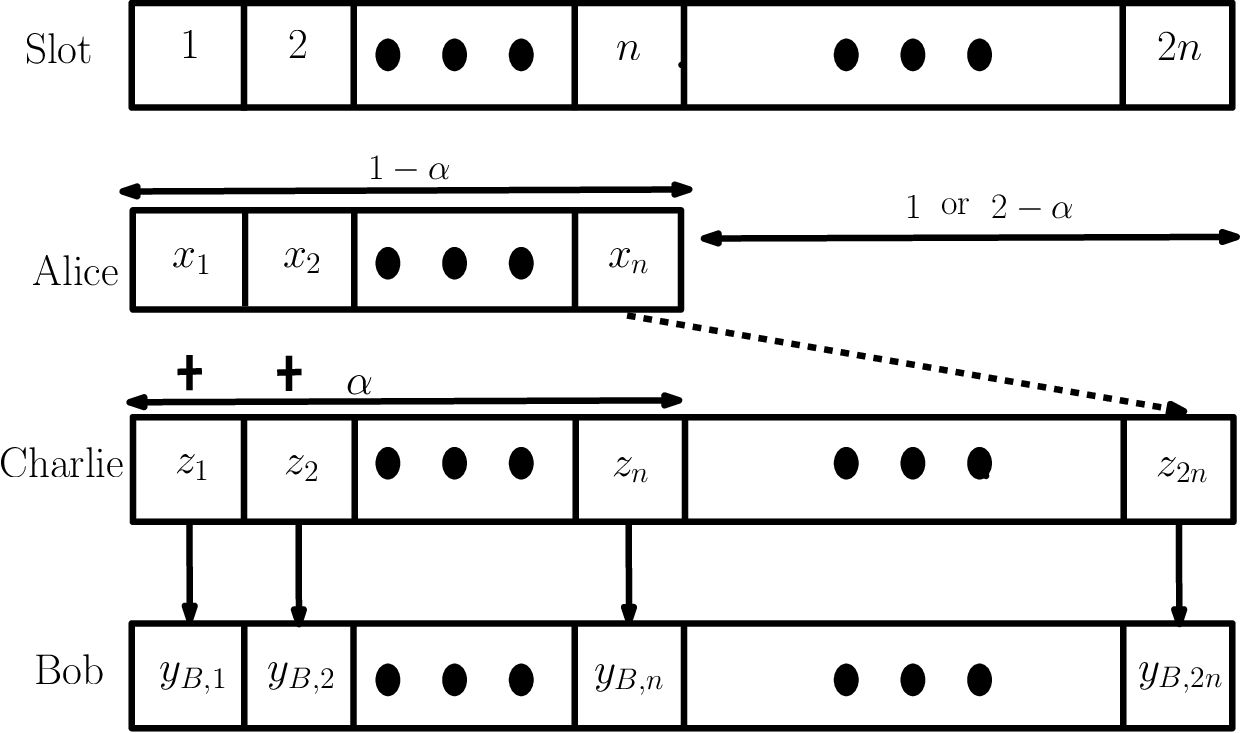}
    \label{fcbdiag}}%
    \hfil
    \subfloat[]
{
\includegraphics[scale=0.27]{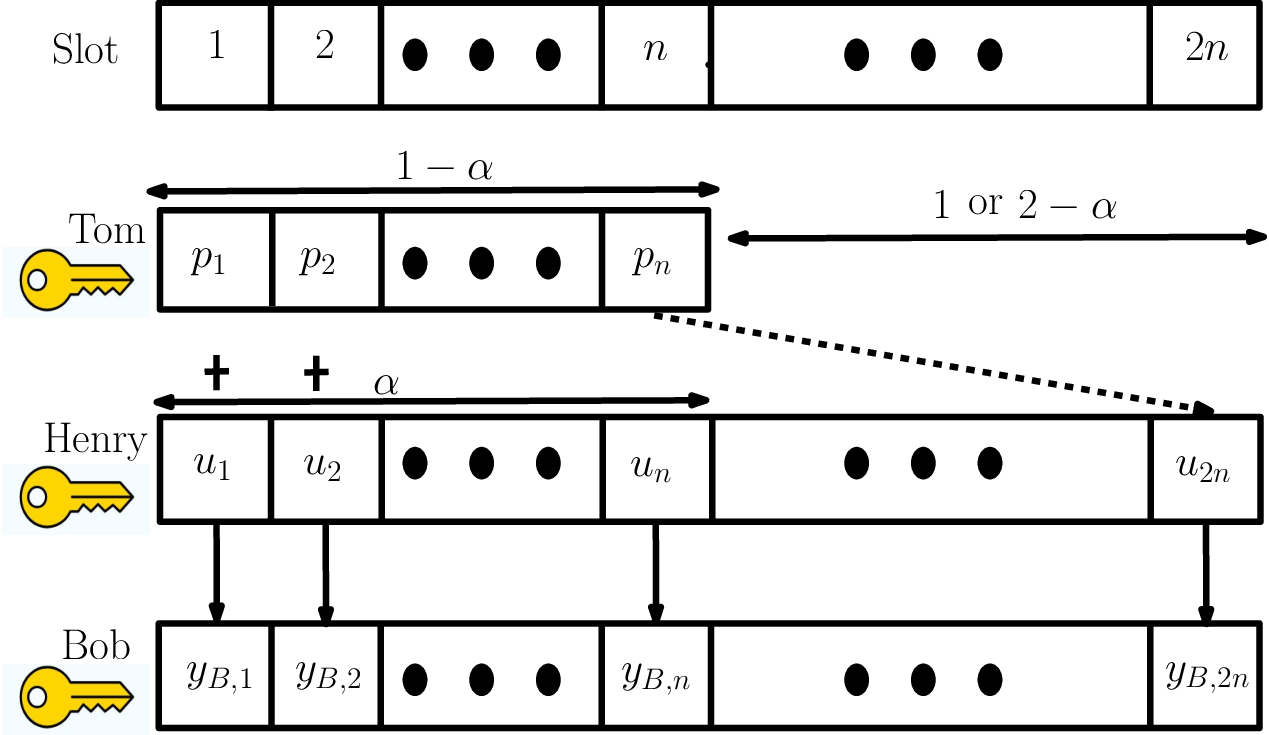}
    \label{fhbdiag}%
\includegraphics[scale=0.27]{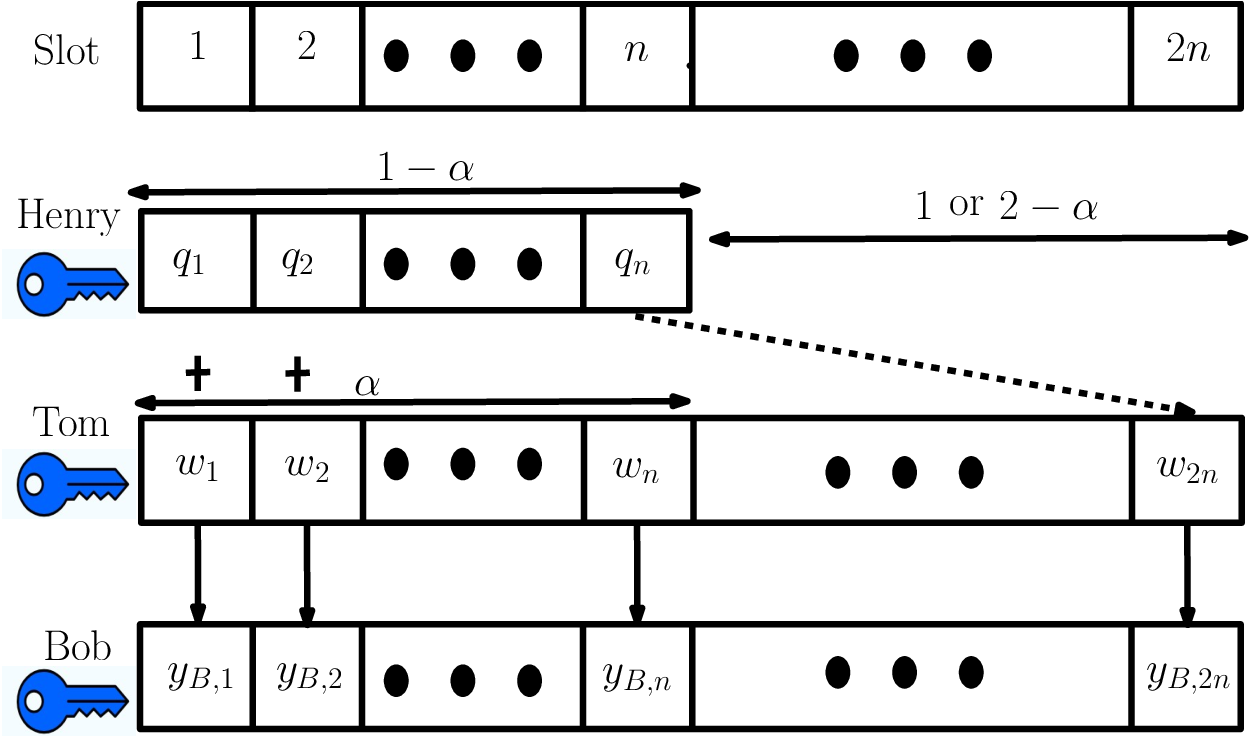}
    \label{fhbdiag}}%
    \caption {The frame structure of information symbols over  (a) the $f_{CB}$ band, and (b) over the $f_{HB}$ and the $f_{TB}$ bands. 
    }
    \label{fcb2diags}
\end{figure*}

The frame structure of information symbols over the $f_{CB}$ band is divided across $2n$ time-slots, as shown in Fig. \ref{fcbdiag}.
During the first $n$ time slots, Alice shifts to the $f_{CB}$ band, wherein Alice and Charlie cooperatively transmit their information symbols to Bob using a portion of their energies.
In the $k^{th}$ time-slot, where $k \in [n]$, Alice uses  OOK and Charlie uses his constellation $\mathcal{S}$, where $\mathcal{S} \subseteq \mathbb{C}$,  
to transmit their information symbols to Bob. In addition to this, Alice and Charlie scale their information symbols by $\sqrt{1-\alpha}$ and $\sqrt{\alpha}$, respectively, where $\alpha \in (0, 1)$ represents the energy-splitting factor. The received symbol at Bob during the $k^{th}$ time-slot, denoted using $y_{B,k}$, is given by
\begin{eqnarray}{rcl}
\hspace{-4mm}{   y_{B,k}=\sqrt{1-\alpha}h_{AB,k}x_k+\sqrt{\alpha}z_{k} h_{CB,k}+n_{B,k},}
\end{eqnarray}
where $h_{AB,k} \sim \mathcal{CN}(0, 1)$ denotes the channel between Alice and Bob, $x_k\in \{0, 1\}$
denotes Alice's bits, 
$z_k \in \mathcal{S}$, where $\mathcal{S}$ denotes the complex constellation used by Charlie (this could be $M$-PSK, $M$-QAM),
 $h_{CB,k}\sim \mathcal{CN}(0, 1)$ is the channel between Charlie and Bob, 
$n_{B,k}\sim \mathcal{CN}(0, N_0)$ is the additive white Gaussian noise (AWGN) at Bob, and the subscript $k$ denotes the $k^{th}$
time-slot. 
As Alice is the victim node, it is imperative to communicate her bits with utmost reliability. As a result, Charlie, who is equipped with an FDR, listens to Alice's transmitted  bit  while transmitting his symbol during the $k^{th}$ time-slot, such that the received vector at Charlie, denoted using $\mathbf{y}_{C,k}$, is given by
\begin{eqnarray}{rcl}
\mathbf{y}_{C,k}=\sqrt{1-\alpha}\mathbf{h}_{AC,k}x_k+ \mathbf{h}_{CC,k}+\mathbf{n}_{C,k},
\end{eqnarray}
where $\mathbf{h}_{AC,k}\sim \mathcal{CN}(\mathbf{0}_{N_C}, \sigma^{2}_{AC}\mathbf{I}_{N_C})$ is the $N_C \times 1$ channel between Alice and Charlie, $\mathbf{h}_{CC,k} \sim \mathcal{CN} (\mathbf{0}_{N_C},\alpha \rho \mathbf{I}_{N_C})$ is  $N_C \times 1$ loop interference (LI) channel at Charlie, and $\mathbf{n}_{C,k}\sim \mathcal{CN}(\mathbf{0}_{N_C},N_0 \mathbf{I}_{N_C})$ is the AWGN at Charlie, $\rho \in (0,1)$ is the LI parameter at Charlie.
Consequently, Charlie decodes Alice's bit transmitted in the $k^{th}$ time-slot, denoted using $\hat{x}_k$, and incorporates this decoded bit into his transmitted information symbol in the form of energy and phase modification during the $(n+k)^{th}$ time-slot.
If  $\hat{x}_k=0$, Charlie adds an additional phase shift of $\pi/M$ and modifies the energy to $2-\alpha$ of his information symbol, such that the received symbol at Bob during the $(n+k) ^{th}$ time-slot is given by
\begin{eqnarray}{rcl}
y_{B,n+k}=\sqrt{2-\alpha}h_{CB,n+k}z_{n+k} e^{
\frac{\iota\pi}{M}}+n_{B,n+k},
\end{eqnarray}
where $h_{CB,n+k}\sim \mathcal{CN}(0, 1)$ is the channel between Charlie and Bob, 
$z_{n+k} \in \mathcal{S}$, $\iota = \sqrt{-1}$,
$n_{B,n+k}\sim \mathcal{CN}(0, N_0)$ is the AWGN at Bob, and the subscript $(n+k)$ denotes the $(n+k)^{th}$ time-slot. Conversely, if $\hat{x}_k=1$, Charlie transmits his symbol without any modification in the energy or phase, such that 
\begin{eqnarray}{rcl}
y_{B,n+k}=h_{CB,n+k}z_{n+k} + n_{B,n+k}.
\end{eqnarray}

\subsection{Strategy over the $f_{AB}$ band}\label{fabtheory}
To force Dave to believe that Alice has not vacated the $f_{AB}$ band, it is imperative to maintain OOK over the $f_{AB}$ band. As a result, Alice and Charlie pour their residual energies of $\alpha$ and $1-\alpha$, respectively, during the $k^{th}$ time-slot using a preshared pseudo-random bit sequence, denoted using $\mathbf{a}$, where $\mathbf{a} =[a_{1}, a_{2}, \ldots, a_{n}]$ and $a_{k}\in \{0,1\}$. The received symbol at Dave during the $k^{th}$ time-slot is of the form $\sqrt{\alpha}h_{AD,k}a_k+\sqrt{1-\alpha}a_k h_{CD,k}+n_{D,k}$,
where $h_{AD,k}\sim \mathcal{CN}(0, 1)$ is the channel between Alice and Dave, $h_{CD,k}\sim \mathcal{CN}(0, 1)$ is the channel between Charlie and Dave,  $n_{D,k}\sim \mathcal{CN}(0, N_0)$ is AWGN at Dave. Subsequently, during the $(n+k)^{th}$ time-slot, Charlie keeps silent, while Alice locally handles the formation of pseudorandom OOK sequence.

\subsection{Strategy over the $f_{HB}$ band}\label{fhbtheory}
Out of the $L-2$ users present in the network other than Alice and Charlie, we ask $L_C$ users to form $\frac{L_C}{2}$ groups of two users each, to mimic the signalling scheme over the $f_{CB}$ band.
As shown in Fig. \ref{fhbdiag}, the users in a pair, use $\alpha$ energy to communicate their information to Bob, on their allocated frequency, and
pour the remaining $1-\alpha$ energy on their partner's frequency band using a secret-key. Note the secret-keys used by both the users in a pair to pour $1-\alpha$ energy in each other's band are different, and these secret-keys are also shared with Bob. 
Let Tom and Henry,  be one such pair who mimic the signalling scheme over the $f_{CB}$ band. In this section, we will only explain the strategy over Henry's frequency band, i.e., the $f_{HB}$ band, owing to the fact that signalling scheme in all the $L_C$ frequency bands have the same structure.
Tom, Henry and Bob have  preshared pseudo-random bit sequence, 
denoted using $\mathbf{p}$, where $\mathbf{p} =[p_{1}, p_{2}, \ldots, p_{n}]$ and $p_{k}\in \{0,1\}$, as shown in the left figure of Fig. \ref{fhbdiag}.
The contents of $\mathbf{p}$ determine if Tom will pour energy over the $f_{HB}$ band in the $k^{th}$ time-slot, and also how Henry will transmit his information symbol in the $(n+k) ^{th}$ time-slot. 
If $p_{k}=1$, Tom pours $1-\alpha$ energy over the $f_{HB}$ band in the $k^{th}$ time-slot, and otherwise he keeps silent, while Henry transmits his information symbol with $\alpha$ energy. 
The received symbol at Bob in $k^{th}$ time-slot over the $f_{HB}$ band, denoted using $y_{B,k}$, is given by
\begin{eqnarray}{rcl}
    y_{B,k}=\sqrt{1-\alpha}h_{TB,k} p_{k}+\sqrt{\alpha}u_{k} h_{HB,k}+n_{B,k},
\end{eqnarray}
where $h_{TB,k}\sim \mathcal{CN}(0, 1)$ is the channel between Tom and Bob, $u_{k} \in \mathcal{S}$,  $h_{HB,k}\sim \mathcal{CN}(0, 1)$ is the channel between Henry and Bob,  $n_{B,k}\sim \mathcal{CN}(0, N_0)$ is the AWGN at Bob.\footnote{For the ease of notations, we use $y_{B,k}$ and $y_{B,n+k}$, to denote the received symbols at Bob, for both the $f_{CB}$ and the $f_{HB}$ bands.} 
Furthermore, if $p_{k}=1$, Henry transmits his information symbol without any modifications during the $(n+k) ^{th}$ time-slot, such that the received symbol at Bob, denoted using  $y_{B, n+k}$, is given by
\begin{eqnarray}{rcl}
y_{B,n+k}=h_{HB,n+k}u_{n+k} + n_{B,n+k},
\end{eqnarray}
where $h_{HB,n+k}\sim \mathcal{CN}(0, 1)$ is the channel between Henry and Bob, 
$u_{n+k}  \in \mathcal{S}$, 
$n_{B,n+k}\sim \mathcal{CN}(0, N_0)$ is the AWGN at Bob. However, if  $p_{k}=0$, Henry adds an additional phase shift of $\pi/M$ and increases the energy to $2-\alpha$ of his information symbol, such that the received symbol at Bob during the $(n+k) ^{th}$ time-slot is given by
\begin{eqnarray}{rcl}
y_{B,n+k}=\sqrt{2-\alpha}h_{HB,n+k}u_{n+k} e^{
\frac{\iota\pi}{M}}+n_{B,n+k}.
\end{eqnarray}

In the second figure of Fig. \ref{fhbdiag}, we present the frame structure of the information symbols over Tom's frequency band, denoted using $f_{TB}$. Here, $q_k$ denotes preshared pseudo-random bit between Tom, Henry and Bob, and $w_{k}$ denotes Tom's information symbol. From the figure, we observe that the signalling scheme of the $f_{TB}$ band is same as that for the $f_{HB}$ band.

 \begin{figure*}[h]
 \vspace{-4mm}
\begin{eqnarray}{rcl}\label{JMAP}
\hat{r}_k, \hat{s}_k, \hat{s}_{n+k} &=& \arg \mathop {\max }\limits_{r_k,s_k,s_{n+k}} f\left( y_{B,k},y_{B,n+k} \left| x_k \right.=r_k, z_k=e^{-\frac{\iota 2 \pi s_k}{M}},  z_{n+k}=e^{-\frac{\iota 2 \pi s_{n+k}}{M}}, h_{CB,k},h_{CB,n+k} \right)
\end{eqnarray}
\vspace{-1mm}
\begin{small}
    \begin{eqnarray}{rcl}\label{Pethcbhb1}
\hspace{-5mm}  Pe^{cb}_{th1}  = \frac{2(C_{11} +C_{13}+ V_{11}+V_{13})+ C_{12} +  V_{12} + C_{14} + V_{14}}{2}, \quad
Pe^{hb}_{th1}  = \frac{2(C_{11} +C_{13}) +C_{12}+ C_{14}}{2}
  \end{eqnarray}
  \vspace{-3mm}
 \begin{eqnarray}{rcl}\label{Pethcbhb2}
Pe^{cb}_{th2}=\frac{1}{2}[ 2 P_{11} (V_{21}+C_{21})+  2 P_{00} (V_{21}+C_{22})+ (P_{01}+P_{10}) (1-V_{21})], \quad Pe^{hb}_{th2}  = \frac{2(C_{21} +C_{22}) +V_{22}+ V_{23}}{2}
  \end{eqnarray}
  \vspace{-3mm}
   \begin{eqnarray}{rcl}\label{CV1}
C_{11}=\sum_{i=1}^3 \frac{k_{i}}{\frac{t_{i}\alpha}{N_{0}}+1}, \quad C_{12}=\sum_{i=1}^3 \frac{k_{i}}{\frac{2t_{i}\alpha}{N_{0}}+1},\quad C_{13}=\sum_{i=1}^3 \frac{k_{i}}{\frac{t_{i}\alpha}{N_{1b}}+1},\quad C_{14}=\sum_{i=1}^3 \frac{k_{i}}{\frac{2t_{i}\alpha}{N_{1b}}+1},\quad
V_{13} =\left(\frac{N_{1b} \sqrt{\alpha}}{(1-\alpha)^2}+1\right)^{-1} 
\end{eqnarray}
\vspace{-2mm}
\begin{eqnarray}{rcl}\label{CV2}
V_{14} =\left(\frac{N_{1b} \sqrt{2\alpha}}{(1-\alpha)^2}+1\right)^{-1},\quad 
V_{11}=\left(\frac{N_0}{N_{1b}}\right)^{\frac{N_{1b}}{1-\alpha}}V_{13},\quad
V_{12} =\left(\frac{N_0}{N_{1b}}\right)^{\frac{N_{1b}}{1-\alpha}}V_{14},\quad
C_{21}=\sum_{i=1}^3 \frac{k_{i}}{\frac{t_{i}}{2N_{0}}+1}
 \end{eqnarray}
 \vspace{-3mm}
 \begin{eqnarray}{rcl}\label{CV3}
C_{22}=\sum_{i=1}^3 \frac{k_{i}}{\frac{t_{i}(2-\alpha)}{N_{0}}+1}, \quad
V_{21}=\sum_{i=1}^3 \frac{k_{i}}{\frac{t_{i}e}{2N_{0}}+1},\quad 
V_{22}=\sum_{i=1}^3 \frac{k_{i}}{\frac{2t_{i}(2-\alpha)}{N_{0}}+1},\quad 
V_{23}=\sum_{i=1}^3 \frac{k_{i}}{\frac{2t_{i}}{N_{0}}+1}
  \end{eqnarray}
\end{small}
  \hrule
\end{figure*}

 \section{Error Analysis of NCMS}

Given that the signalling method over the $f_{CB}$ and the $f_{HB}$ bands are different, Bob employs different decoding strategies to decode the transmitted symbols over these frequency bands.
We will first discuss the decoding strategy used at Bob to retrieve the transmitted symbols over the $f_{CB}$ band, followed by the  decoding strategy at Bob to retrieve the transmitted symbols over the $f_{HB}$ band, assuming that all the users (other than Alice), use $M$-ary phase-shift keying ($M$-PSK) to communicate with Bob.

\subsection{Decoding strategy for symbols transmitted over $f_{CB}$ band}
Recall that, to provide reliability to Alice's information bit, we ask Charlie to embed Alice's information $x_k$ into his transmitted symbol during the $(n+k)^{th}$ time-slot (i.e. $z_{n+k}$), in the form of energy and phase modifications. 
As a result, we conclude that the decoding strategy used at Charlie will ultimately influence the decoding strategy at Bob.
First, we will discuss the decoding strategy used at Charlie to decode $x_k$.
On the basis of the received vector  $\mathbf{y}_{C,k}$, Charlie decodes Alice's information bit ($\hat{x}_k$) using non-coherent detection, after setting an optimal threshold $\tau$. To achieve this, Charlie computes the energy of  $\mathbf{y}_{C,k}$, denoted using $|\mathbf{y}_{C,k}|^2$,  and compares it with $\tau$. If $|\mathbf{y}_{C,k}|^2 >\tau$, then $\hat{x}_k =1$, otherwise, $\hat{x}_k=0$. For this decision rule, the crossover probabilities, i.e. the probabilities associated with decoding  bit-$0$ as bit-$1$, and bit-$1$ as bit-$0$, are represented using $P_{01}$ and $P_{10}$, respectively. As stated earlier, the decoding strategy at Bob depends on the decoding strategy at Charlie, as a result, the values of $P_{01}$ and $P_{10}$ are preshared with Bob using a secret-key.

Now we will discuss the decoding strategy at Bob used to decode the transmitted symbols over the $f_{CB}$ band. 
To jointly decode Alice's and Charlie's information symbols transmitted during  $k^{th}$ and $(n+k)^{th}$ time-slot, Bob uses Joint Maximum A Posteriori (JMAP) decoder, such that the decoding metric is given by \eqref{JMAP}, where
$f\left( y_{B,k},y_{B,n+k} \left| x_k, z_k,  z_{n+k}, h_{CB,k},h_{CB,n+k}\right. \right)$
denotes the conditional probability density function (CPDF) associated with $y_{B,k}$ and $y_{B,n+k}$, given $x_k$, $z_k$, $z_{n+k}$, $ h_{CB,k}$ and $h_{CB,n+k}$.
Here, $r_k \in \{ 0,1\}$ and $s_k, s_{n+k} \in \{0, 1, \ldots, M-1\}$ denote the search space of the JMAP decoder. Although the decoding metric given in \eqref{JMAP} is optimal, it has the following shortcomings. First, the CPDF given in \eqref{JMAP} consists of Gaussian mixtures, which are scaled by the probabilities of decoding error at Charlie, and the difficulties in using Gaussian mixtures are well known in the literature.
Second, owing to the fact that the JMAP decoder given in \eqref{JMAP} simultaneously decodes three symbols ($x_k$, $z_k$ and $z_{n+k}$), 
the decoding complexity is $\mathcal{O} (2M^2)$, which is not desired.
Based on the above-mentioned shortcomings, we conclude that deriving the expression of the probability of decoding error at Bob
using the JMAP decoder proposed in \eqref{JMAP} is intractable. Consequently, we propose a decoder, called Disjoint Decoder, whose decoding metric does not contain any Gaussian mixture, and the decoding complexity is $\mathcal{O} (2M)$. As the name suggests, in this decoder Bob decodes the information symbols transmitted during the $k^{th}$ and the $(n+k)^{th}$ time-slot  independent of each other. Based on the received symbol $y_{B,k}$ during the $k^{th}$ time-slot, Bob decodes Charlie's symbol $z_k$. 
Subsequently, based on $y_{B,n+k}$ during the $(n+k)^{th}$ time-slot, Bob decodes $x_k$ and $z_{n+k}$. 
Now, we will discuss the decoding strategy at Bob to decode 
$z_k$, followed by the decoding strategy for $x_k$ and $z_{n+k}$. 

To decode $z_k$ from $y_{B,k}$, the decoding metric at Bob is
 \begin{eqnarray}{rcl}\label{JMAP1}
\hat{s}_k &=& \arg \mathop {\max }\limits_{s_k} 
f\left( y_{B,k}| x_{k}, z_{k}=e^{-\frac{\iota 2 \pi s_k}{M}}, h_{CB,k} \right),
\end{eqnarray}
where $f\left( y_{B,k}| x_{k}, z_{k}, h_{CB,k} \right)$ denotes the CPDF of $y_{B,k}$ given $x_{k}$, $z_{k}$ and  $h_{CB,k}$.
Similarly, the decoding metric at Bob to decode $z_{n+k}$ from $y_{B,n+k}$ is given by
\begin{eqnarray}{rcl}\label{JMAP2}
\hat{r}_{k},\hat{s}_{n+k} &=& \arg \mathop {\max }\limits_{r_k,s_{n+k}} 
\!f\left(y_{B,n+k}| x_{k}=r_{k}, \right.
\nonumber \\ & &\left. z_{n+k}=\!e^{-\frac{\iota 2 \pi s_{n+k}}{M}}, h_{CB,n+k} \right),
\end{eqnarray}  

\noindent where $f\left( y_{B,n+k}| x_{k}, z_{n+k}, h_{CB,n+k} \right)$ denotes the CPDF of $y_{B,n+k}$ given $x_{k}$, $z_{n+k}$ and  $h_{CB,n+k}$.

\subsection{Decoding strategy for symbols transmitted over $f_{HB}$ band}
Recall that based on ${p}_k$, Tom pours $1-\alpha$ energy in the $k^{th}$ time-slot, and Henry  changes the phase and the energy of his symbol $u_{n+k}$ in the $(n+k)^{th}$ time-slot.  
As a result, we conclude the received symbol at Bob during the $k^{th}$ and the $(n+k)^{th}$, i.e., $y_{B,k}$ and $y_{B,n+k}$, respectively, are independent of each other, given the knowledge of $p_k$ at Bob. 
Consequently, based on $y_{B,k}$, Bob decodes $u_{k}$ in the $k^{th}$ time-slot, and based on  $y_{B,n+k}$, Bob decodes $u_{n+k}$ in the $(n+k)^{th}$ time-slot. Note the decoding complexity at Bob to decode the symbols transmitted over $f_{HB}$ is $\mathcal{O} (M)$. Now, we will explain the decoding strategy at Bob to decode  $u_{k}$, followed by the decoding strategy for $u_{n+k}$. 

To decode $u_{k}$ from $y_{B,k}$, Bob uses MAP decoder, such that the decoding metric of the MAP decoder is given by 
 \begin{eqnarray}{rcl}\label{MAP1}
\hat{u}_k &=& \arg \mathop {\max }\limits_{u_k} 
f\left( y_{B,k}| p_k, u_{k}, h_{HB,k} \right),
\end{eqnarray}
where $f\left( y_{B,k}| p_k, u_{k}, h_{HB,k} \right)$ denotes the CPDF of $y_{B,k}$ given $p_k$, $u_{k}$ and  $h_{HB,k}$, and $u_k$ denote the search space of the MAP decoder. 
Similarly, to 
decode $u_{n+k}$ from $y_{B,n+k}$, Bob uses MAP decoder, such that the decoding metric is given by 
     \begin{eqnarray}{rcl}\label{MAP2}
\hat{u}_{n+k} &=& \arg \mathop {\max }\limits_{u_{n+k}} 
f\left(\! y_{B,n+k}| p_k, u_{n+k}, h_{HB,n+k} \right)\!,
\end{eqnarray}
\noindent where $f\left( y_{B,n+k}| p_k, u_{n+k}, h_{HB,n+k} \right)$ denotes the CPDF of $y_{B,n+k}$ given $p_k$, $u_{n+k}$ and  $h_{HB,n+k}$, and $u_{n+k}$ denote the search space of the MAP decoder.

\subsection{Overall probability of decoding error at Bob}

Let $Pe_{AC}$  and $Pe_{H}$ denote the  average probability of decoding error at Bob associated with the $f_{CB}$ band and the $f_{HB}$ band, respectively, (averaged over all the channel realisations). The expression of $Pe_{AC}$ can be obtained by computing the probability of various error events from \eqref{JMAP},  and similarly, the expression of $Pe_{H}$ can be obtained by computing the probability of various error events from \eqref{MAP1} and \eqref{MAP2}. Also, let $Pe_{NH}$ denote the probability of decoding error at Bob associated with the users not mimicking the transmissions over the $f_{CB}$ band, which are $L-L_C-2$ in number, and it can be obtained using first principles. Therefore, the average probability of decoding error at Bob averaged over all the network users, denoted using $Pe$, is given by
\begin{eqnarray}{rcl}\label{Peall}
  Pe = \frac{Pe_{AC}+L_C Pe_{H} + (L-L_C-2)Pe_{NH}}{L}. 
\end{eqnarray}

\begin{theorem}\label{ThPe}
    An upper bound on $Pe$, denoted by $Pe_{th}$, is

    \begin{small}
           \begin{eqnarray}{rcl}\label{Peth}
Pe_{th} & = &\frac{Pe^{cb}_{th1}\!+\!Pe^{cb}_{th2}\!+\! L_C(Pe^{hb}_{th1}\!+\!Pe^{hb}_{th2})\!+\! 2(L\!-\!L_C-\!2)Pe^{nh}_{th}}{L}. \nonumber \\
\end{eqnarray}       
    \end{small}

\noindent    Here, $Pe^{cb}_{th1}$, $Pe^{cb}_{th2}$, $Pe^{hb}_{th1}$ and $Pe^{hb}_{th2}$ denote the probability of decoding error using the decoding metric given in \eqref{JMAP1}, \eqref{JMAP2}, \eqref{MAP1} and \eqref{MAP2}, respectively, and $P_{00}=1-P_{01}$ and $P_{11}=1-P_{10}$. Also, $Pe^{nh}_{th}$ denotes the upper bound on probability of decoding error at Bob associated with those users who are not mimicking the transmissions over the $f_{CB}$ band, which are $L-L_C-2$ in number. The expression of $Pe^{cb}_{th1}$ and $Pe^{hb}_{th1}$ are given in \eqref{Pethcbhb1},  and $Pe^{cb}_{th2}$ and  $Pe^{hb}_{th2}$ are given in \eqref{Pethcbhb2}, and the expression of $Pe^{nh}_{th}$ can be derived using first principles. Also, the expressions of the variables in $Pe^{cb}_{th1}$, $Pe^{hb}_{th1}$, $Pe^{cb}_{th2}$ and    $Pe^{hb}_{th2}$  are given in \eqref{CV1}, \eqref{CV2} and \eqref{CV3}, where $k_{1} = 0.168$, $k_{2} = 0.144$, $k_{3} = 0.002$, $t_{1} = 0.876$, $t_{2} = 0.525$, $t_{3} = 0.603$,  $N_{1b}=N_0+1-\alpha$ and $e=\sqrt{3-\alpha-2\sqrt{2-\alpha} cos (\pi/M)}$.
                                            
\end{theorem}

\subsection{Simulation Results}
\begin{figure}[ht!]
   \begin{center}
       {\includegraphics[scale=0.20]{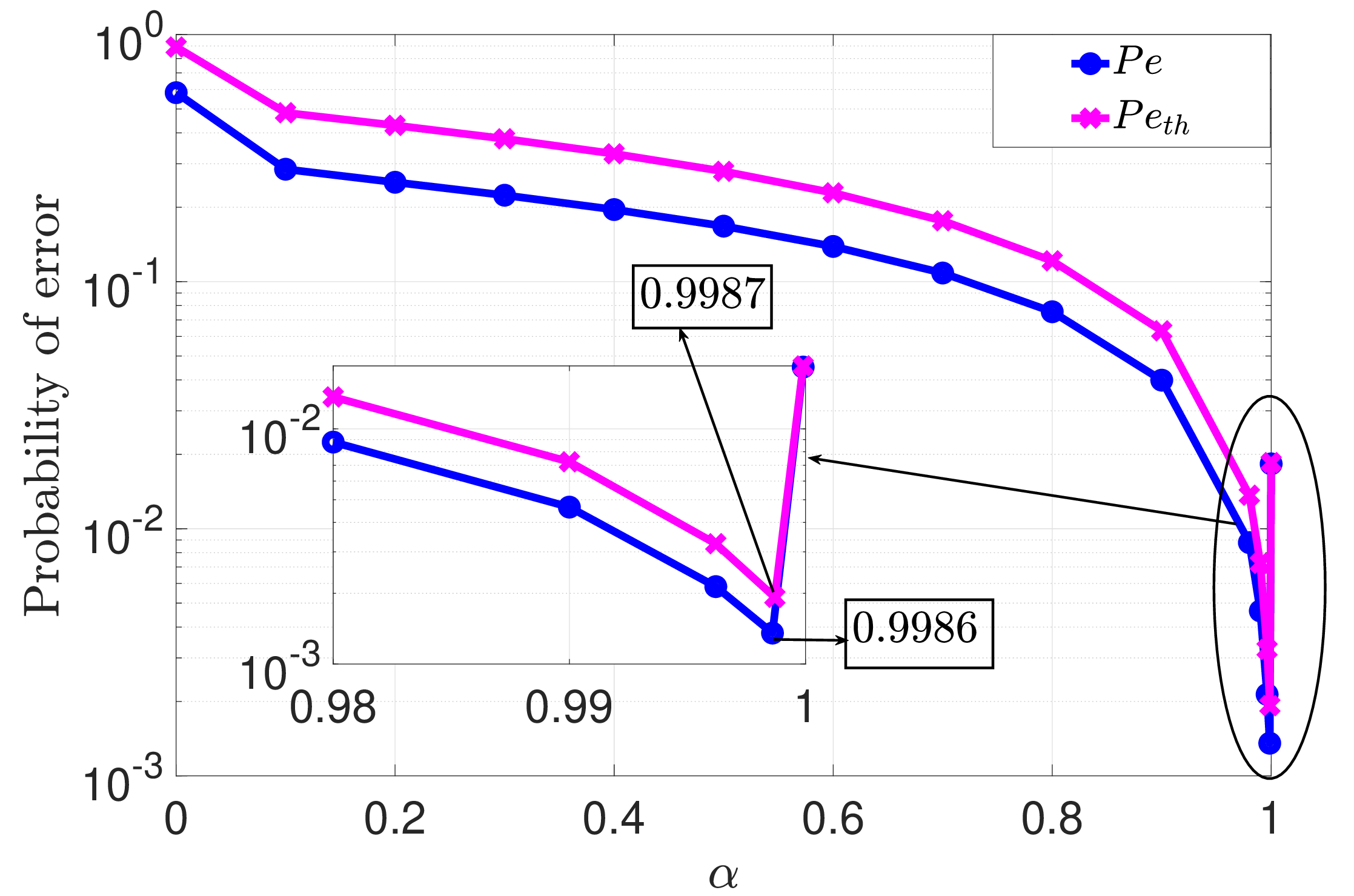}}
   \end{center}  
   \vspace{-0.35cm}
\caption{Plots depicting $Pe$ and $Pe_{th}$, as a function of $\alpha$, for $L_C=10$, $N_C=4$, and at an SNR of $35$ dB.  We observe that the minima of $Pe$ and $Pe_{th}$, are around the same value of $\alpha$.}
\label{Pethsim}
    \end{figure}

To prove the correctness of the derived upper bound $Pe_{th}$, we fix $L_C$ and $L$, and then study the behaviour of $Pe_{th}$ and $Pe$, as a function of $\alpha$.
Owing to the fact that average probability of decoding error at Bob associated with the users not mimicking the signalling scheme over the $f_{CB}$ band does not depend on $\alpha$,  for this analysis, we will only consider the average probability of decoding error associated with the $f_{CB}$ and the $f_{HB}$ bands, which are $L_C$ in number. In Fig. \ref{Pethsim}, we plot $Pe_{th}$ as a function of $\alpha$, marked using magenta, for $L_C=10$, $N_C=4$, and at an SNR of $35$ dB. In the same figure, for the same set of parameters, we use monte-carlo simulations to plot $Pe$ as a function of $\alpha$, marked using blue. From these plots, we have the following  observations. First, the curve of $Pe_{th}$ sits slightly above the curve of $Pe$, which validates the fact that  $Pe_{th}$ is  an upper bound on $Pe$. Second, we observe that the curve of  $Pe_{th}$ and $Pe$ follow a similar trend, and their minima are around the same value of $\alpha$.

    \begin{figure}[ht!]
   \begin{center}
       {\includegraphics[scale=0.20]{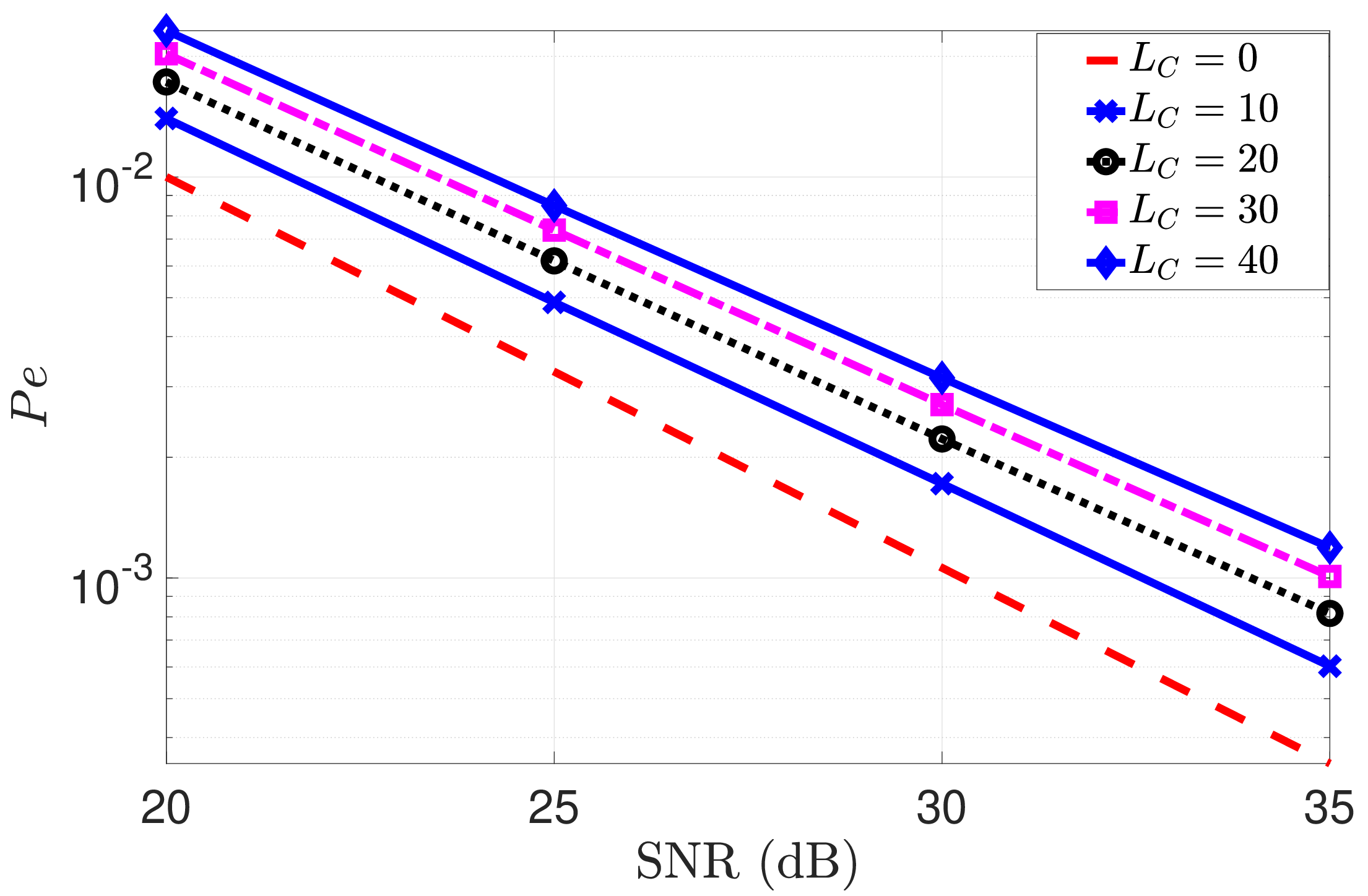}}
   \end{center}  
   \vspace{-0.35cm}
\caption{Plot depicting $Pe$ as a function of SNR, for different values of $L_C$ and $L=42$. From these plots, we observe that as $L_C$ increases, $Pe$ increases.}
\label{PeSNR}
    \end{figure}
Next, we fix $L$ and vary $L_C$, and study the behaviour of $Pe$ as a function of SNR. In this regard,
in Fig. \ref{PeSNR}, we use monte-carlo simulations to plot $Pe$ as a function of SNR, for $L=42$, and different values of $L_C$. To generate these plots, for a given $L_C$ and SNR, we compute the operating value of $\alpha$ such that $Pe_{th}$ is minimised, and substituted these values in \eqref{Peall}, to get $Pe$.  From these plots, we have the following observations. First, for a given $L_C$, $Pe$ decreases with SNR, which is required from any reliable communication system. Second, we observe that for a given SNR, $Pe$ increases with $L_C$. This is because $Pe_{H}>Pe_{NH}$, this is a penalty for mimicking the signalling scheme over $f_{CB}$ band, and as $L_C$ increases, $(L-L_C-2)$ decreases. Stitching these two events, we conclude that as $L_C$ increases, $L_C Pe_{H}$ increases, while $(L-L_C-2) Pe_{NH}$ decreases, consequently, $Pe$ increases.

\begin{figure}[ht!]
   \begin{center}
       {\includegraphics[scale=0.18]{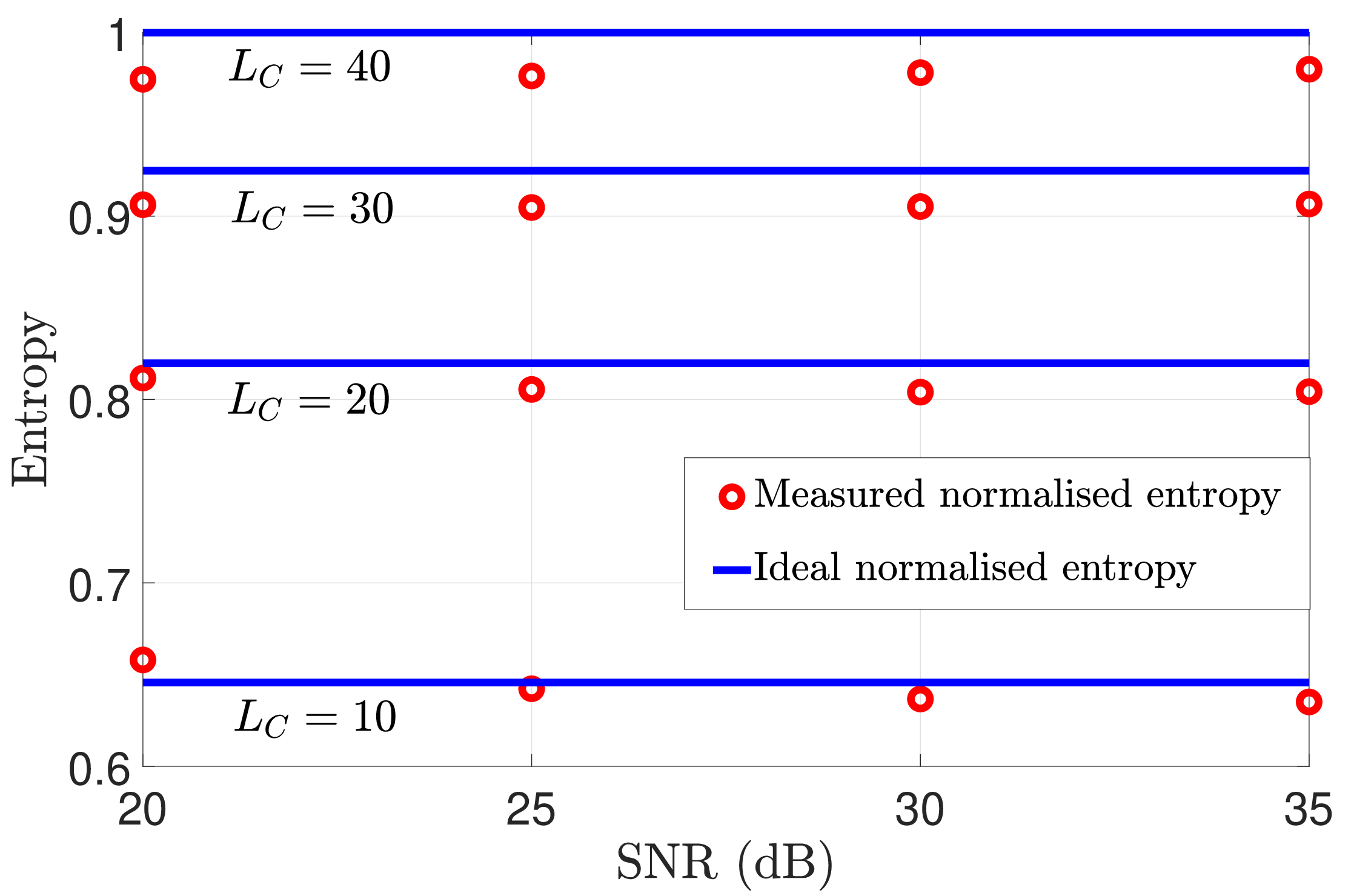}}
   \end{center}  
   \vspace{-0.35cm}
\caption{Figure shows the plot of normalised entropy as a function of SNR, for different values of $L_C$ and $L=42$. From these plots, we observe that as $L_C$ increases, the normalised entropy increases. }
\label{KP_Case1}
    \end{figure}

\section{Covertness Analysis of NCMS}
Recall that Kerchoffs' principle states that all the details about the signalling scheme should be made public, except the secret-keys. In this regard, Dave has the following knowledge about the scheme discussed in Section \ref{CH}.
First, Dave knows that Alice has moved to an another band, to communicate her information reliably to Bob. Alice and  the incumbent user of that band, pour $1-\alpha$ and $\alpha$ energies, respectively,  to communicate their information symbols to Bob in the $k^{th}$ time-slot.
Also, that user is equipped with an FDR, as a result, to provide an added layer of reliability to Alice's information bits, that user incorporates Alice's bits into its symbols in the form of phase and energy modifications in the $(n+k)^{th}$ time-slot.\footnote{Note that the probability of decoding error at that incumbent user associated with Alice's bits are shared with Bob using a secret-key, as a result, Dave does not know the values of $P_{00}$, $P_{11}$, $P_{01}$ and $P_{10}$.}
Second, Dave knows that Alice and the incumbent user of that band, pour $\alpha$ and $1-\alpha$ energies, respectively, over $f_{AB}$ band to form pseudo-random bit sequences, using a shared secret-key.
Third, Dave knows the operating value of energy-splitting factor $\alpha$.
We highlight that although Dave knows all the details of NCMS, he does not know  the band to which Alice has moved to communicate her information to Bob. As a result, his objective is to find out the helper's frequency band using the knowledge of the countermeasure.

To find out the helper's frequency band, Dave tunes into all network frequencies (except Alice's band), collects the received symbols for a frame-length of $f$ symbols. 
The received symbol at Dave for a frequency band $l$ in a given slot of frame, in the $k^{th}$ and the $(n+k)^{th}$ time-slot, are denoted using $d_{k,l}$ and $d_{n+k,l}$, respectively, where $l \in \{1, \ldots, L-1 \}$. 
Next, Dave uses two sets of CPDF, one containing the CPDFs on countermeasure setup, and the other containing the CPDF for no countermeasure setup.  The CPDF of the countermeasure setup for $x_k=0$ and $x_k=1$, are denoted using $D_1$ and $D_2$, respectively, and the CPDF  for no countermeasure setup is denoted using $D_3$. 
Next, using $d_{k,l}$, $d_{n+k,l}$ and $\alpha$, Dave counts the number of times $D_1$ or $D_2$ gets maximised averaged over $f$ for a frequency $l$, which is denoted using $m_l$. Dave repeats this operation for all the $L-1$ network frequencies. Note that the higher the value of $m_l$, the higher the probability that the band $l$ is the helper's frequency band. As a result, based on $m_l$,  Dave assigns scores to all the network frequencies, denoted using $P_l$, by using the softmax function
$ P_l= exp(d \, m_l),  
$
where $d$ is the scaling factor. 
Using  $P_l$, the entropy associated with the helper's frequency band, denoted using $H$, is given by 
\begin{eqnarray}{rcl}
 H= -\sum_{l=1}^{L-1}  P_{n,l}  \log_2( P_{n,l}),
\end{eqnarray}
where $ P_{n,l}$ denotes the probability, which is given by $ P_{n,l}=\frac{P_l}{\sum_{l=1}^{L-1}P_l}$. The measured normalised entropy, denoted using $H_{norm}$, is given by $H_{norm}=\frac{H}{\log_2(L-1)}$, where $H_{norm} \in (0,1)$, such that $H_{norm}=0$ and $H_{norm}=1$, correspond to no uncertainty and maximum uncertainty, respectively, regarding the helper's frequency band at Dave.

\subsection{Simulation results}    
To show the security strength of NCMS in maintaining uncertainty at Dave regarding the helper's frequency band, in Fig. \ref{KP_Case1}, we  use monte-carlo simulations to plot the measured normalised entropy $H_{norm}$, for various combinations of $L_C$ and SNR, $d=10$, $f=200$ and $L=42$, along with the ideal normalised entropy, which is given by $\frac{\log_2(L_C+1)}{\log_2(L-1)}$, marked using horizontal line. Note that for a given $L_C$, the operating value of $\alpha$ is chosen in such a way that $Pe_{th}$ is minimum.
From these plots, we observe that $H_{norm}$ is close to the ideal normalised entropy, which shows the security strength of NCMS against Kerchoffs' principle. Also, we observe that
as $L_C$ increases, $H_{norm}$ increases. This is due to the fact that as more users mimic the transmissions over the $f_{CB}$ band, the difference in the value of normalised assigned probability $P_{n,l}$ minimises, consequently, $H_{norm}$ increases.

\begin{table*}[h]\caption{Solutions to Problem 1 and Problem \ref{ProbTh} for $L=42$.}\label{diff_alpha}
{%
\begin{center}
\begin{scriptsize}
\begin{tabular}{|l|l|l|l|l|l|l|l|l|l|}
\hline
  & \multicolumn{2}{c|}{$\delta=0.658$} & \multicolumn{2}{c|}{$\delta=0.8117$} & \multicolumn{2}{c|}{$\delta=0.9062$} & \multicolumn{2}{c|}{$\delta=0.9746$} \\ \hline
\!\text{SNR}\! & $(\alpha^{*}, L_C^\ast)$  & $(\alpha^{\dagger}, L_C^\dagger)$ & $(\alpha^{*}, L_C^\ast)$  & $(\alpha^{\dagger}, L_C^\dagger)$ & $(\alpha^{*}, L_C^\ast)$  & $(\alpha^{\dagger}, L_C^\dagger)$ & $(\alpha^{*}, L_C^\ast)$  & $(\alpha^{\dagger}, L_C^\dagger)$  \\ \hline
\!$30$\!           
& \!$(0.9970,10)$\!           & \!$(0.9973,10)$\!     
& \!$(0.9978,20)$\!          & \!$(0.9978,20)$\!     
& \!$(0.9978,30)$\!          & \!$(0.9980,28)$\! 
& \!$(0.9981,40)$\!           & \!$(0.9982,38)$\!\\ \hline
  & \multicolumn{2}{c|}{$\delta=0.6351$} & \multicolumn{2}{c|}{$\delta=0.8043$} & \multicolumn{2}{c|}{$\delta=0.9066$} & \multicolumn{2}{c|}{$\delta=0.9801$} \\ \hline
\!$35$\!          
& \!$(0.9986,10)$\!           & \!$(0.9987,10)$\!     
& \!$(0.9988,20)$\!           & \!$(0.9990,20)$ \!    
& \!$(0.9990,30)$\!           & \!$(0.9990,28)$\! 
& \!$(0.9991,40)$\!           & \!$(0.9991,38)$\!\\ \hline
\end{tabular}
\end{scriptsize}
\end{center}
}
\end{table*}

Based on the discussion in this section and the preceding section, we conclude that as $L_C$ increases, the reliability at Bob averaged over all users degrades, while the entropy at Dave regarding the helper’s band improves. This suggests that there is a tradeoff between reliability and entropy, and  $L_C$ should be chosen such that the reliability is maximum, subject to the measured entropy is within an acceptable range.

\section{Parameter Optimization of NCMS}
To communicate both reliably and covertly, we present an optimization problem in Problem \ref{Probsim}, which provides the optimal value of $\alpha$ and $L_C$, denoted using $\alpha^{*}$ and $L_C^\ast$, respectively. This problem is posed such that the average probability of decoding error at Bob $Pe$ is minimised, with the measured normalised entropy $H_{norm}$ lower bounded by $\delta$, for some $0\leq \delta \leq 1$.

  \begin{mdframed}
\begin{problem}
\label{Probsim}
For a given $L$, $N_C$ and SNR, we solve: 
 \begin{eqnarray}{rcl}\label{Probsimexp}
L_C^\ast,\alpha^{*}= \arg \mathop {\min }\limits_{\alpha \in (0, 1),L_C \in \{1,\ldots,L-1 \}} Pe,
  \end{eqnarray}
subject to: $H_{norm}\geq \delta$.

\end{problem}
\end{mdframed}

\begin{figure}[ht!]
   \begin{center}
       {\includegraphics[scale=0.20]{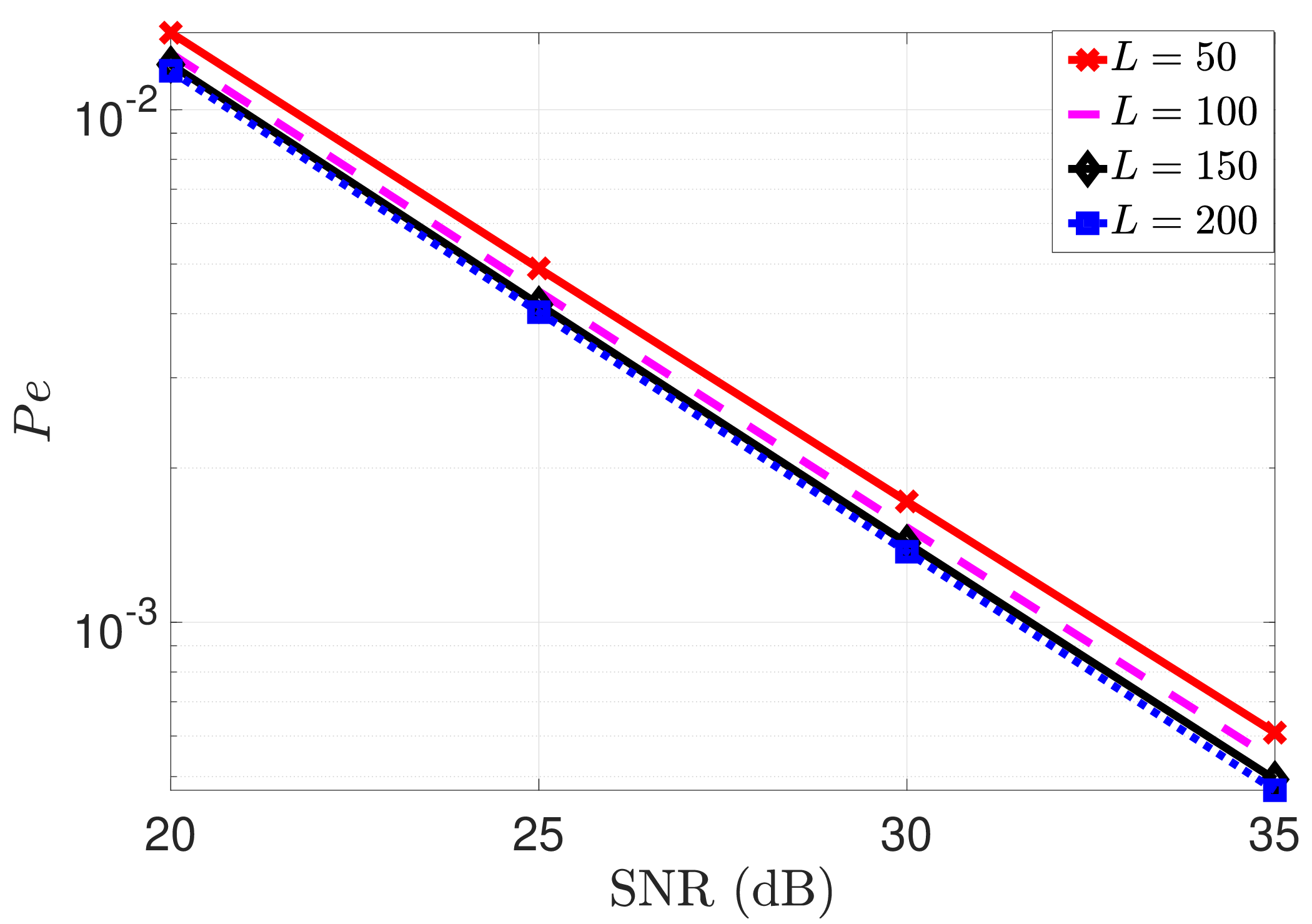}}
   \end{center} 
   \vspace{-0.35cm}
\caption{Plots depicting $Pe$, as a function of SNR, for $\delta=0.7$ and different values of $L$. To generate these plots we use Table \ref{diff_alpha_2}.}
\label{PeSNR_L}
    \end{figure}

However, given that we have an upper bound on  $Pe$ (i.e., $Pe_{th}$, given in Theorem \ref{ThPe}), instead of the true expression, and there is no expression of measured normalised entropy $H_{norm}$, we propose to solve Problem \ref{ProbTh}, which provides $L_C^\dagger$ and $\alpha^{\dagger}$, such that $\frac{\log_2(L_C+1)}{\log_2(L-1)} \geq \delta$, and compare it with the solution provided by Problem \ref{Probsim}. 
In Table \ref{diff_alpha}, we tabulate the solutions to Problem \ref{Probsim} and Problem \ref{ProbTh}, for different values of SNR and  $\delta$. 
From the table, we observe that the solutions to Problem \ref{Probsim} and Problem \ref{ProbTh} are close to each other, which suggests the accuracy of modified optimization problem. 
From the table, we observe that for a fix $L$ and SNR, as $\delta$ increases, $L_C^{\ast}$ (or $L_C^{\dagger}$) increases, consequently $Pe$ increases. This observation is in agreement with the plots shown in Fig. \ref{PeSNR}. In Table \ref{diff_alpha_2}, we present the solutions to Problem \ref{ProbTh}, for $\delta=0.7$, and different values of $L$. 
From the table, we observe that for a fix $\delta$, as $L$ increases, $L_C^\dagger$ increases. Now, as the difference between $L$ and $L_C^\dagger$ increases, we observe that the  contribution of $(L-L_C^\dagger-2)Pe_{NH}$ to $Pe$ becomes dominant (see \eqref{Peall}), as a result, $Pe$ decreases with $L$, which is shown in Fig. \ref{PeSNR_L}. Thus, as $L$ increases, the overall error performance of the network improves for a given $\delta$.
\begin{mdframed}
\begin{problem}
\label{ProbTh}
For a given $L$, $N_C$ and SNR, we solve: 
 \begin{eqnarray}{rcl}\label{ProbThexp}
L_C^{\dagger},\alpha^{\dagger}= \arg \mathop {\min }\limits_{\alpha \in (0, 1),L_C \in \{1,\ldots,L-1 \}} Pe_{th},
  \end{eqnarray}
  subject to:  $\frac{\log_2(L_C+1)}{\log_2(L-1)} \geq \delta$.
\end{problem}
\end{mdframed}

\begin{table}[h]\caption{Solutions to Problem \ref{ProbTh} for different values of $L$}\label{diff_alpha_2}
{%
\begin{center}
\begin{scriptsize}
\begin{tabular}{|l|l|l|l|l|l|}
\hline
\multirow{3}{*}{SNR}& $L=50$   & $L=100$  & $L=150$  & $L=200$  \\ \cline{2-5}
  & $L_C^{\dagger}=14$   & $L_C^{\dagger}=24$  & $L_C^{\dagger}=32$  & $L_C^{\dagger}=40$  \\ \cline{2-5}
    & $\alpha^{\dagger}$   & $\alpha^{\dagger}$  & $\alpha^{\dagger}$  & $\alpha^{\dagger}$  \\ \hline
\!$20$\!      
& \!$0.9907$         & \!$0.9928$  
&\!$0.9938$          & \!$0.9946$\!\\ \hline
\!$25$\!              
& \!$0.9950$         & \!$0.9959$
& \!$0.9963$          & \!$0.9967$\\ \hline
\!$30$\!               
& \!$0.9975$         & \!$0.9979$
& \!$0.9981$        & \!$0.9983$\\ \hline
\!$35$\!              
& \!$0.9988$       & \!$0.9990$ 
& \!$0.9991$       & \!$0.9991$\\ \hline
\end{tabular}
\end{scriptsize}
\end{center}
}
\end{table}

\section{Communication Overheads and Future Work}

In this section we highlight various overheads associated with NCMS from implementation aspects. Firstly, NCMS is applicable when there is no strict latency constraints on the messages of the victim node. This is because the base station takes some time-slots to make transition from the normal setup to the countermeasure setup, after realising that the victim's band is jammed. For instance, to implement NCMS, the base-station (i) has to identify the list of users willing to participate in NCMS, (ii) form groups of two users based on their geographical locations and synchronization capabilities, and finally (iii) distribute the necessary power allocation parameters to all the participating users. Besides these overheads associated with migrating to NCMS mode, the participating users must be provided secret-keys to synchronously pour energies into their partner's frequency band. In particular, to implement NCMS, users in each pair need a key-rate of 0.5 bits per channel use of common randomness, which accounts to an overall key-rate of $\frac{L_{c}}{2}$ bits per channel use in the network. Given that this is the first work of its kind to propose a countermeasure against an ISAJ adversary abiding Kerchoffs' principle, we believe there are interesting research directions for future work especially addressing the problems of communication and secret-key overheads.

\end{document}